\documentclass[aps,prd,twocolumn,superscriptaddress,groupedaddress,nofootinbib]{revtex4-1}  % for review and submission

\usepackage[utf8]{inputenc}

\usepackage{graphicx}  % needed for figures
\usepackage{dcolumn}   % needed for some tables\usepackage{bm}        % for math
\usepackage{amssymb}   % for math
\usepackage{appendix}

\newcommand{\e}{\mbox{e}}
\newcommand{\cp}{{\cal P}}
\newcommand{\lzmz}{\lambda_{\mbox{\tiny ZMZ}}}

\begin{document}

% The following information is for internal review, please remove them for submission
\widetext

% the following line is for submission, including submission to the arXiv!!
%\hspace{5.2in} \mbox{Fermilab-Pub-04/xxx-E}

\title{Ideal topological gas in the high temperature phase of SU(3) gauge
  theory}
   % D0 authors (remove the first 3 lines
                             % of this file prior to submission, they
                             % contain a time stamp for the authorlist)
                             % (includes institutions and visitors)

\author{ Réka Á. Vig}

\affiliation{University of Debrecen, H-4032 Debrecen, Bem tér 18/A, Hungary\\}

\author{Tamás G. Kovács} 

\affiliation{Eötvös Loránd University, H-1117 Budapest, Pázmány Péter sétány
  1/A, Hungary\\ and \\
 Institute for Nuclear Research, H-4026 Debrecen, Bem tér 18/c, Hungary}

\date{\today}
\begin{abstract}
We show that the nature of the topological fluctuations in $SU(3)$ gauge
theory changes drastically at the finite temperature phase
transition. Starting from temperatures right above the phase transition
topological fluctuations come in well separated lumps of unit charge that form
a non-interacting ideal gas. Our analysis is based on a novel method to count
not only the net topological charge, but also separately the number of
positively and negatively charged lumps in lattice configurations using the
spectrum of the overlap Dirac operator. This enables us to determine the joint
distribution of the number of positively and negatively charged topological
objects, and we find this distribution to be consistent with that of an ideal
gas of unit charged topological objects.
\end{abstract}

\maketitle

%\section{Introduction}
%   \label{sec:introduction}

The presence of topologically nontrivial gauge field configurations is a
peculiar feature of QCD that has important phenomenological consequences. Most
recently this was highlighted by calculations to estimate the axion mass
\cite{Borsanyi:2016ksw}-\cite{Petreczky:2016vrs}. An essential ingredient of
the calculation was the determination of the temperature-dependence of the
topological susceptibility up to temperatures well above the QCD crossover
temperature to the quark-gluon plasma.

At very high temperatures, fluctuations of the topological charge are strongly
suppressed and occur in the form of localized lumps of action, carrying
topological charge $\pm 1$. These objects are probably close in their shape
and other properties to solutions of the classical Euclidean gauge field
equations, i.e.\ instantons, or rather their finite temperature counterparts,
calorons\footnote{Since these objects are not exact solutions of the field
  equations, they are not exactly calorons or anticalorons, it would be more
  appropriate to call them {\it topological objects}. Nevertheless, for
  simplicity we will mostly use the word {\it instanton} or {\it caloron} for
  them.} \cite{Kraan:1998pm}-\cite{Gattringer:2002tg}. Moreover, since at high
temperatures, fluctuations of the topological charge are strongly suppressed,
calorons (and anticalorons) are expected to form a dilute gas and their size
is limited by the inverse temperature.

These properties of the caloron gas motivate the so called dilute instanton
gas approximation (DIGA) which --in principle-- makes it possible to
calculate the temperature-dependence of the topological susceptibility
perturbatively. However, the topological susceptibility determined in lattice
simulations differs by an order of magnitude from the DIGA predictions even at
temperatures as high as $5-10T_c$
\cite{Borsanyi:2016ksw}-\cite{Petreczky:2016vrs}. It is thus clear that at
least one of the two assumptions that the DIGA is based on is not
satisfied. These two assumptions, both of which are expected to be valid at
high enough temperatures are: {\it {\bf A1:} The probability of one instanton
  occurring in a given volume can be calculated perturbatively in the
  semiclassical approximation.}  {\it {\bf A2:} The instantons gas is so
  dilute that interactions among (anti)instantons can be neglected, the gas of
  topological objects is an ideal gas. }

Assumption {\it A1} has been recently reconsidered, but despite the correction
of the previously grossly underestimated uncertainty of the semiclassical
one-instanton calculation, there is still at least a $3\sigma$ discrepancy
between the lattice and the DIGA result for the topological susceptibility
\cite{Boccaletti:2020mxu}.  In the present paper we focus on assumption {\it
  A2} and study interactions among topological objects in the quenched
approximation of QCD, just above the finite temperature phase transition. Even
apart from the axion problem, a full determination of instanton interactions
just above $T_c$ is interesting in itself, as it can shed light on how typical
gauge field configurations change as the system crosses into the high
temperature phase. This might also help us better understand the chiral and
deconfining transition in QCD with dynamical quarks. 

To see how interactions among instantons could be detected, our starting point
is a non-interacting instanton gas. In a free, non-interacting topological gas
the number distribution of topological objects can be characterized with a
single parameter, the topological susceptibility $\chi = \frac{\langle Q^2
  \rangle}{V}$, where $Q=n_i-n_a$ is the topological charge (the number of
instantons minus the number of antiinstantons), $V$ is the space-time volume
and $\langle . \rangle$ denotes the expectation with respect to the path
integral. In an ideal topological gas all higher cumulants of the distribution
can be exactly calculated in terms of $\chi$. Any deviations from these
ideal-gas cumulants are a result of interactions among topological objects.

In the recent literature several quenched lattice calculations of the lowest
non-trivial cumulant\footnote{We note that in the literature usually the value
  of $b_2=-B_2/12$ is quoted, as that is the coefficient appearing in the
  expansion of the partition function in terms of the theta parameter.}
\begin{equation}
  B_2 =  \frac{\langle Q^4 \rangle - 3 \langle Q^2 \rangle^2}{\langle Q^2
    \rangle} 
\end{equation}
appeared \cite{Borsanyi:2015cka,Bonati:2013tt}. The most precise calculation
reports that even though above $1.15T_c$ the value of $B_2$ is consistent with
$1$, its ideal-gas value, just above the phase transition, at
$1.045T_c$ it is still 1.27(7) \cite{Bonati:2013tt}.

For a more complete assessment of the situation, more information would be
desirable about the distribution of the number of topological objects, beyond
the first nontrivial cumulant of $Q$. However, higher cumulants of the
distribution are notoriously hard to calculate, and even the full topological
charge distribution can in principle miss subtle correlations among instantons
and antiinstantons. Full information about that is contained only in the joint
distribution of the number of instantons and antiinstantons. 

The problem is that while there are well-established methods to compute the
topological charge $Q=n_i-n_a$ in lattice simulations, there is no easy way to
determine $n_i$ and $n_a$ separately in lattice configurations\footnote{One
  possibility would be to analyze the structure of the topological charge
  density and locate individual lumps in it. However, this would be rather
  cumbersome and would be plagued by uncertainties.}.  In the present paper we
introduce a novel method to compute $n_i$ and $n_a$ separately, and determine
their joint distribution. Our method is based on the low-lying spectrum of the
overlap Dirac operator. In particular, our main observation is that mixing
instanton-antiinstanton zero modes constitute a distinct part of the Dirac
spectrum close to zero, and can be reliably separated from the rest of the
spectrum. Counting the number of these close to zero modes together with the
exact zero modes of the overlap operator provides a way to determine not only
the topological charge $Q$, but also the total number of topological objects
$n_i+n_a$.

Besides yielding much more information than just the cumulant $b_2$, our
method has another advantage compared to previous studies. By construction it
always gives integer numbers, and thereby avoids the ambiguities that plague
the definitions of the topological charge based on gauge field operators. In
that case, the values of the topological charge are not integers and need to
be multiplicatively renormalized, and -- if the full charge distribution is
needed -- also rounded to integers. Higher moments of the distribution, like
$b_2$ are very sensitive to these ambiguities.

For the present study we used quenched lattice configurations generated at
$T=1.045T_c$ on lattices of temporal extension $N_t=8$ and aspect ratio 3 and
4. For both spatial volumes we determined $n_i$ and $n_a$ on 5k lattice
configurations. In the smaller volume, we found that the number distribution
of topological objects significantly deviated from the expectation based on a
free non-interacting gas. In contrast, the larger volume showed no such
deviation. Although in finite temperature lattice QCD an aspect ratio of 3 is
usually considered safe in terms of finite (spatial) volume corrections, we
show here that the unexpectedly large finite volume corrections are due to the
proximity of the phase transition. We conclude that in quenched QCD already
slightly above $T_c$ the number distribution of topological objects is
consistent with that of a gas of free topological objects. 

Let us first motivate the main tool used in our study, the separation of the
bulk of the spectrum and the topology-related close to zero modes. It is known
that in the presence of an isolated instanton (or antiinstanton) the Euclidean
Dirac operator has an exact zero mode with chirality $+1$ ($-1$)
\cite{Atiyah:rm}. In the field of a well separated instanton and antiinstanton
the two would be zero eigenvalues split slightly and produce two complex
conjugate eigenvalues. The splitting is controled by the spatial distance of
the topological objects (relative to their size), as well as their orientation
in group space. Generally the farther away the two objects are, the smaller
the splitting is, and in the limit of infinite separation the splitting tends
to zero \cite{Schafer:1996wv}. In this way a dilute gas of topological objects
is expected to produce not only $|Q|=|n_i-n_a|$ exact zero modes,
corresponding to the net topological charge, but also $n_i+n_a-|Q|$ small
Dirac eigenvalues, from the mixing of opposite chirality instanton and
antiinstanton would be zero modes. Motivated by the instanton liquid model, we
call the region in the spectrum containing these modes the Zero Mode Zone
(ZMZ).

 It should be already clear from the above discussion that as the temperature
 gets higher and topological fluctuations become sparser, the near zero modes
 of topological origin will be closer to the origin. At the same time, the low
 end of the bulk of the spectrum, the lowest non-topological modes, controlled
 by the Matsubara frequency, will move higher as the temperature
 increases. Therefore, at high enough temperature the ZMZ should be well
 separated from the bulk of the spectrum. In what follows, we will demonstrate
 that already slightly above the finite temperature phase transition the ZMZ
 can be reliably separated from the rest of the Dirac spectrum, provided a
 chirally symmetric Dirac operator, such as the overlap
 \cite{Narayanan:1993ss} is used.

Already in the early days of the overlap it was noticed that above $T_c$,
besides the expected exact zero modes, the overlap Dirac spectrum also
contains an unexpectedly large number of very small close to zero eigenvalues
\cite{Edwards:1999zm}. This enhancement of the low end of the Dirac spectrum
resulted in a spike in the spectral density, well separated from the bulk of
the spectrum. This came as a surprise, since above $T_c$ the restoration of
chiral symmetry would imply a vanishing spectral density at zero virtuality,
due to the Banks-Casher relation. The spike in the spectral density was
conjectured to contain mixing would be zero modes of instantons and
antiinstantons. Subsequent work confirmed that this enhancement of the
spectral density is neither a discretization, nor a quenched artifact
\cite{Alexandru:2015fxa}. More recently the appearance of this spike in the
Dirac spectrum was speculated to signal a genuinely new state of strongly
interacting matter, intermediate between the low temperature hadronic and the
high temperature quark-gluon plasma state \cite{Alexandru:2019gdm}.

In the present work we analyze the statistical properties of the eigenvalues
in this spike of the spectral density in a high statistics quenched $SU(3)$
lattice study. We show that the statistics of these eigenvalues is to a high
precision consistent with the assumption that they are produced by mixing
instanton and antiinstanton would be zero modes. To this end we use quenched
gauge ensembles generated with the Wilson gauge action at $\beta=6.09$ and
temporal lattice extension $N_t=8$. This corresponds to a temperature of
$T=1.045T_c$, just above the finite temperature transition that in the
quenched $SU(3)$ case is a first order phase transition. For the detailed
statistical analysis we used two ensembles of gauge configurations with
spatial extension $L=24$ and $32$, both containing 5000 configurations. In
addition, to check finite volume effects in the spectral density and the
Polyakov loop distribution, we also had an ensemble of 600 configurations on a
larger spatial volume $L^3=40^3$. The negative mass parameter of the overlap
Wilson kernel was set to $M=-1.3$, and two steps of hex smearing
\cite{Capitani:2006ni} were performed on the gauge links before inserting them
into the Wilson kernel.  The statistical analysis we report here was performed
on the overlap Dirac eigenvalues of smallest magnitude with $|\lambda|/T_c <
2.0$ for all ensembles.
 
Since we want to compare the statistics of small Dirac eigenvalues with that
of non-interacting topological objects, we first summarize our expectations in
such an ideal gas of topological objects. In the absence of any interaction
among them, both the number of instantons $n_i$ and that of antiinstantons
$n_a$ are expected to follow independent and identical Poisson distributions
with mean $V \chi/2$ proportional to the volume, where $V$ is the four-volume
of the system and $\chi$ will turn out to be the topological
susceptibility. The joint distribution
\begin{equation}
  P(n_i,n_a) = \e^{-V\chi} \frac{(V\chi/2)^{n_i+n_a}}{n_i! n_a!} 
\end{equation}
of $n_i$ and $n_a$ can be used to compute all the relevant physical quantities
of this free topological gas in terms of the single parameter $\chi$. In
particular the topological susceptibility is
\begin{eqnarray}
  \frac{1}{V} \langle Q^2 \rangle & = &
  \sum_{n_i=0}^\infty \sum_{n_a=0}^\infty P(n_i,n_a) (n_i-n_a)^2 = \nonumber \\
  & = &  \overline{n_i^2}  + \overline{n_a^2} -
  2 \overline{n_i} \ \overline{n_a} = \chi,  
\end{eqnarray}
where expectations like $\overline{n_i}$ are understood to be with respect to
the respective Poisson distribution. The topological charge distribution for
$Q \geq 0$ is
\begin{equation}
  P(Q) = \sum_{n=0}^\infty \e^{-V\chi} \frac{(V\chi/2)^{Q+2n}}{(Q+n)! n!} =
  \e^{-V\chi} I_Q(V\chi),
    \label{eq:qdist}
\end{equation}
where $I_Q$ are the Bessel functions of imaginary argument. Due to
time-reversal symmetry the distribution is symmetric, $P(Q)=P(-Q)$.

Another interesting quantity to consider is the distribution of the
total number of topological objects $n=n_i+n_a$,
\begin{equation}
  \cp(n) = \sum_{n_i=0}^n \e^{-V\chi} \frac{(V\chi/2)^n}{n_i! (n-n_i)!} =
    \e^{-V\chi} \frac{ (V\chi)^n}{n!},
    \label{eq:ntdist}
\end{equation}
which is simply a Poisson distribution with mean $V \chi$. 

\begin{figure}
\centering
\includegraphics[width=1\columnwidth]{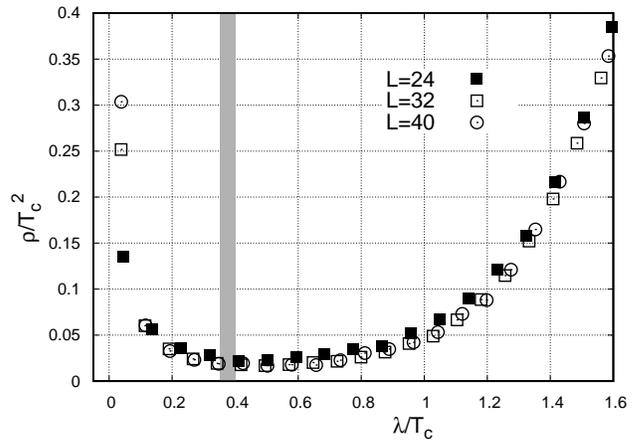}
\caption{\label{fig:specdens} The spectral density of the overlap Dirac
  operator on quenched $SU(3)$ gauge ensembles just above the phase
  transition, at $T=1.045T_c$. The shaded region indicates
  $\lzmz/T_c$, the boundary of the Zero Mode Zone. Eigenmodes below
  this point in the spectrum are related to mixing topological would be zero
  modes.}
\end{figure}

Let us now confront the lattice data with these expectations. In
Fig.\ \ref{fig:specdens} we show the spectral density of the overlap Dirac
operator on the previously mentioned lattice ensembles. Although in the
spontaneously broken phase of the pure gauge theory that we consider here, the
three Polyakov loop sectors are equivalent, the spectrum of the Dirac operator
and the pattern of chiral symmetry breaking/restoration is known to be
strongly dependent on the Polyakov-loop sector \cite{Gattringer:2002tg},
\cite{Chandrasekharan:1995gt}-\cite{Stephanov:1996he}. Since in the present
work we want to study features that are expected to be at least qualitatively
present in QCD with dynamical quarks, here we restrict the analysis to
configurations in the {\it physical} $\mbox{Re} P >0$ Polyakov loop
sector. This is the only one that would be present if dynamical fermions were
to be included. In fact, even the slightest explicit breaking of the $Z(3)$
symmetry by dynamical fermions with arbitrarily large, but finite mass would
force the system into the real Polyakov-loop sector.

The enhancement of the spectral density near zero is clearly seen in
Fig.\ \ref{fig:specdens}.  We note that the exact zero eigenvalues are not
shown here, they would appear as a delta function exactly at zero.

\begin{figure}
\centering
\includegraphics[width=1\columnwidth]{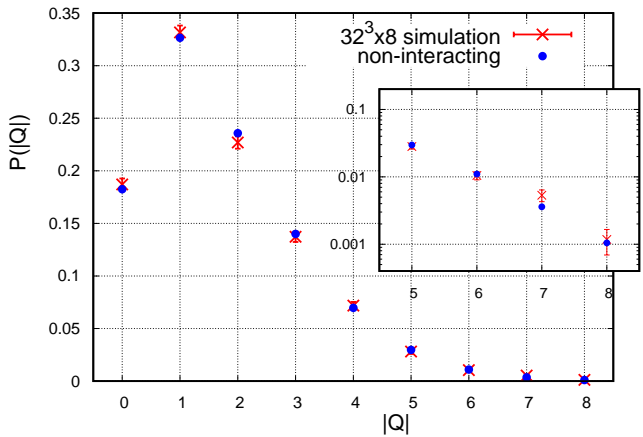}
\caption{\label{fig:qdist} The distribution of the topological charge in our
  lattice simulations and the distribution expected in a free topological gas
  with the same susceptibility. As the distribution is expected to be
  symmetric, positive and negative charges of the same magnitude are counted
  together. The inset shows the tail of the distribution on logarithmic scale.}
\end{figure}

Counting the number of zero eigenvalues allows us to compute the topological
susceptibility $\chi$, as well as the distribution of the topological
charge. In Fig.\ \ref{fig:qdist} we compare the distribution obtained in the
lattice simulation with the one expected in a free topological gas with
susceptibility $\chi$. This is essentially a one-parameter fit of the function
in Eq.\ (\ref{eq:qdist}), the fit parameter being $V\chi$ and the chi squared
per degree of freedom of the fit turns out to be 0.85.

Encouraged by the good agreement between the lattice data and the free
topological gas, we assume that the exact zero modes and the small Dirac
eigenvalues, up to a point $\lzmz$ in the spectrum, are the eigenvalues
associated to the topological objects. In this way, by counting them we count
the number of topological objects $n$ present in the gauge field. To make this
picture consistent, we have to choose $\lzmz$ such that $\langle n \rangle =V
\chi$, as predicted by Eq.\ (\ref{eq:ntdist}) for an ideal topological
gas. Requiring this, results in $\lzmz a=0.045(6)$\footnote{The quoted
  uncertainty includes only the statistical error computed with the
  bootstrap.}, which turns out to be in the depleted region of the spectral
density, separating the spike at zero from the bulk (see
Fig.\ \ref{fig:specdens}). This shows that the ZMZ is indeed well separated
from the bulk of the spectrum.

\begin{figure}
\centering
\includegraphics[width=1\columnwidth]{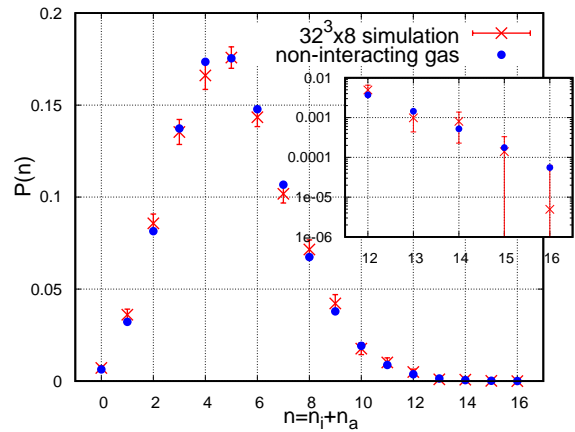}
\caption{The distribution of the number of topological objects computed from
  the number of eigenvalues in the zero mode zone. The inset shows the tail of
  the distribution on logarithmic scale.
  \label{fig:ntdist}}
\end{figure}

We can now identify the total number of eigenvalues in the zero mode zone, the
ones with $|\lambda| < \lzmz$ (including the exact zero modes) with $n$, the
number of topological objects. Counting the eigenvalues in the ZMZ
configuration by configuration, we obtain the distribution of $n$ and in
Fig.\ \ref{fig:ntdist} we compare it with the one expected in a gas of
noninteracting topological objects, given by Eq.\ (\ref{eq:ntdist}). We
emphasize that at this point no fitting is involved, since the only parameter
of this distribution, $V \chi$, had already been determined independently from
the charge distribution. We do not find a significant deviation from the free
topological gas distribution, as the chi squared per degree of freedom of the
deviation is 0.62.

We emphasize that the fact that the distribution of the number of eigenmodes
of magnitude smaller than $\lzmz$ exactly reproduces the distribution we
expect from eigenmodes linked to free topological objects is rather
non-trivial. By choosing $\lzmz$ in the above described manner, we only fixed
the expectation of the distribution, and it follows the expected analytical
form over three orders of magnitude, the whole range where we have data. This
shows not only that the topological objects are indeed non-interacting, but
also that already at this temperature the zero mode zone can be reliably
separated from the bulk.

\begin{figure}
\centering
\includegraphics[width=1\columnwidth]{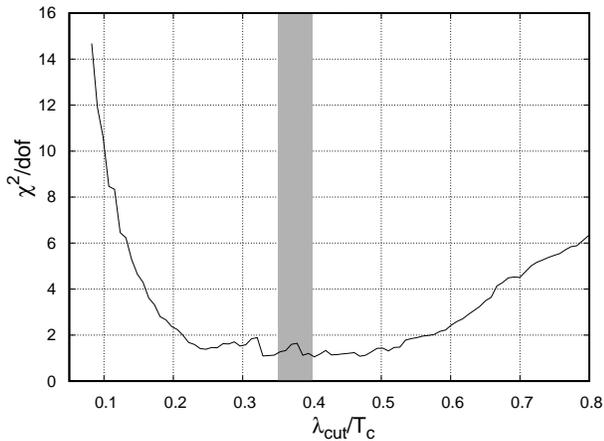}
\caption{The chi squared per degree of freedom of fits to Poisson
  distributions of the distribution of the number of eigenvalues with
  $|\lambda|<\lambda_{cut}$ versus $\lambda_{cut}$. The shaded region shows
  $\lzmz$ with its uncertainty.
  \label{fig:cut_vs_chi}}
\end{figure}

So far we determined $\lzmz$ from the requirement that the total number of
eigenvalues below $\lzmz$ (including the zero modes) be consistent with the
topological susceptibility obtained by counting the zero modes only. But is
$\lzmz$ really a special point in the spectrum? To answer this question we
chose different cuts in the spectrum and checked how close the distribution of
the number of eigenvalues below these cuts are to a Poisson distributions. To
this end we determined the distribution of the number of eigenvalues as a
function of the cut and for each value of the cut plot the chi squared per
degree of freedom of the deviation of the best fit Poisson distribution from
the data. In Fig.\ \ref{fig:cut_vs_chi} we show the results. It can be clearly
seen that there is a range of cuts $0.3< \lambda_{cut}/T_c < 0.6$, where the
distribution is compatible with Poisson, and the previously {\it
  independently} determined $\lzmz$ happens to be in the middle of this range.
This further supports that the valley of the spectral density containing
$\lzmz$ indeed separates the topological modes from the bulk of the
spectrum. This valley, however, is rather wide and even though the spectrum
there is quite depleted and not many eigenvalues are contained there, the
question arises as to how sharply $\lzmz$ is defined within the valley. Given
the present data set, this question cannot be unambiguously answered. It is
possible that with much larger statistics the chi squared test presented in
Fig.~\ref{fig:cut_vs_chi} would further limit the acceptable range for
$\lzmz$. Another possibility is that in the continuum limit the spectrum could
become more depleted in the valley, even a gap could appear there. In that
case the exact location of $\lzmz$ within the gap would not be
important. Finally, it is also possible that even in the continuum limit and
with arbitrarily large statistics there would still be some small ambiguity in
separating the topological modes and the bulk. To explore these possibilities
further would require more extensive simulations that are out of the scope of
the present work.

The finite temperature $SU(3)$ transition is a first order phase transition,
so the correlation length does not diverge, however, large finite volume
corrections cannot be excluded. To assess their importance, we repeated the
analysis in a {\it smaller} volume with linear extension $L=24$. In that
case we found significant deviations from the expected free topological gas
behavior. The resulting chi squared per degree of freedom was 1.99 and 6.29 in
the case of the charge distribution and the distribution of the total number
of topological objects, respectively.

\begin{figure}
\centering
\includegraphics[width=1\columnwidth]{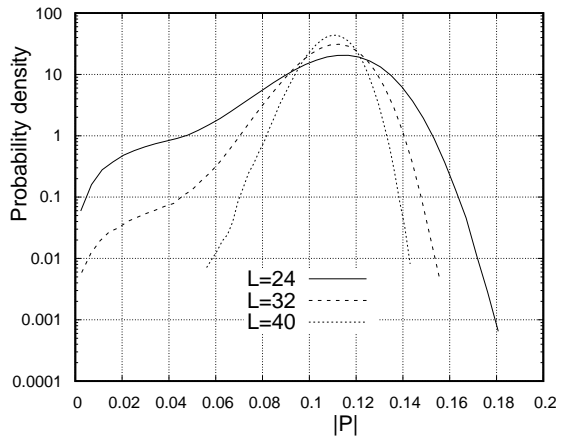}
\caption{The probability distribution of the Polyakov loop for three different
  spatial volumes with linear size $L=24,32$ and $40$. 
  \label{fig:plphist}}
\end{figure}

To understand finite volume corrections in the vicinity of a phase transition,
it is instructive to look at the volume dependence of the distribution of the
order parameter. In Fig.\ \ref{fig:plphist} we show the probability
distribution of the magnitude of the Polyakov loop, the order parameter of the
quenched finite temperature transition. Besides the widening of the
distribution, expected for smaller volumes, the data for $L=24,32$ lattices
also show an unusual enhancement of smaller values of the Polyakov loop. The
reason for this is that in the high temperature phase the $Z(3)$ center
symmetry is spontaneously broken and the system randomly chooses one of the
three $Z(3)$ sectors. However, in a finite volume tunneling among the sectors
is still possible, the tunneling probability is enhanced in smaller
volumes, and configurations in the process of tunneling have small magnitudes
of the Polyakov loop.

\begin{figure}
\centering
\includegraphics[width=1\columnwidth]{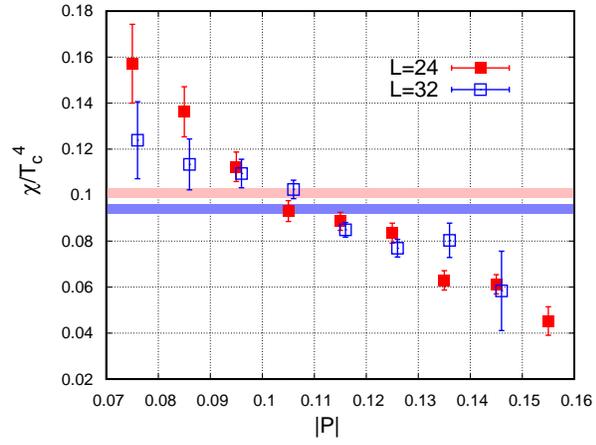}
\caption{The dependence of the topological susceptibility on the Polyakov loop
  for two different spatial volumes with linear size $L=24,32$. The two
  horizontal stripes indicate the susceptibility (with its uncertainty)
  computed for the full ensembles without restricting the Polyakov loop. The
  lower value, indicated with the darker band corresponds to the larger volume.
  \label{fig:chi_vs_plp}}
\end{figure}

As also seen in Fig.\ \ref{fig:plphist}, in larger volumes these tunneling
states get suppressed, however, in smaller volumes they can still give
significant contributions to physical quantities, resulting in large
finite-size effects. To see, how these tunneling states can affect the
topological charge, we looked at the correlation between topology and the
Polyakov loop. The simplest quantity to study is the topological
susceptibility. We computed its dependence on the Polyakov loop by restricting
the averaging of $Q^2$ to configurations with Polyakov loop magnitudes in
intervals of length 0.01. The results for the $L=24$ and 32 ensembles, shown
in Fig.\ \ref{fig:chi_vs_plp}, reveal a strong dependence of the susceptibility
on the Polyakov loop. The previously seen enhanced contribution of the
tunneling region (small Polyakov loop), where the susceptibility is larger,
will thus result in significantly larger topological susceptibilities in
smaller volumes. To have a feeling about the relative importance of the
enhanced region, we note that the probability that $|P|<0.08$ is 0.11, 0.028
and 0.004 for the lattices with linear spatial size size $L=24,32$ and 40
respectively. For the two ensembles shown in Fig.\ \ref{fig:chi_vs_plp} we
also indicated the overall susceptibilities and their uncertainties with the
horizontal stripes. Since on the $L=24$ lattices even the susceptibility
suffers large finite-size effects, it is not surprising that the same occurs
for the distribution of the topological charge and the number of topological
objects.

We would also like to comment on the apparent discrepancy between our results
and those of Ref.\ \cite{Bonati:2013tt} who found a significant $(27(7)\%)$
deviation of the $B_2$ coefficient from its value (1.0) expected in a
non-interacting instanton gas. In fact, our data yields $B_2=1.35(41)$, which
is compatible with that of the above reference. However, judging from their
much smaller uncertainty, their statistics could be more than an order of
magnitude larger than ours. Since it is based on the overlap spectrum, our
method is computationally much more expensive, and in the present study we
could not compete in the statistics, but our method has two advantages. First,
it necessarily yields integer charges and avoids the large ambiguity in $B_2$
due to any small random fluctuations of the topological charge around
integers and the possibly necessary normalization and rounding of the
charges. Secondly, our method allows for a full determination of the joint
distribution of instantons and antiinstantons. It would be interesting to
repeat our calculation using a much larger statistics and to see if there is
any deviation in this distribution from that of the number of noninteracting
instantons.

In the present paper we used a novel way to compute the joint distribution of
the number of topological objects in lattice simulations. We showed that right
above the critical temperature of pure $SU(3)$ gauge theory the distribution
is consistent with the one expected in an ideal gas of non-interacting charges
of unit magnitude. It is remarkable that while below the phase transition
topological fluctuations form a dense medium without easily identifiable
individual lumps \cite{Horvath:2001ir}, right above the phase transition an
ideal gas of well separated topological lumps emerges. Our result also implies
that the most likely explanation of the large discrepancy between the lattice
and DIGA based calculation of the topological susceptibility is that the
topological lumps we found are not close enough in shape to ideal calorons to
warrant a semiclassical treatment.

We expect that --at least on a qualitative level-- this picture of the
topological fluctuations that we found in the quenched case carries over to
QCD with dynamical quarks. However, the fermion determinant might introduce
some interaction even among well separated topological lumps, but to study
that one would need to use a chiral Dirac operator also for the simulation of
the sea quarks. 

%\begin{acknowledgments}
{\em Acknowledgments} TGK was partially supported by the Hungarian National
Research, Development and Innovation Office - NKFIH grant KKP126769. TGK
thanks Matteo Giordano, Sándor Katz and Dániel Nógrádi for helpful
discussions.
%\end{acknowledgments}

\end{document}